\documentclass[9pt, conference]{IEEEtran}
\IEEEoverridecommandlockouts

\usepackage{cite}
\usepackage{amsmath,amssymb,amsfonts}
\usepackage{algorithmic}
\usepackage{graphicx}
\usepackage{textcomp}
\usepackage{xcolor}
\usepackage{rotating}
\usepackage{epstopdf}
\usepackage{bm}
\usepackage{float}
\usepackage{url}

\def\BibTeX{{\rm B\kern-.05em{\sc i\kern-.025em b}\kern-.08em
    T\kern-.1667em\lower.7ex\hbox{E}\kern-.125emX}}

\makeatletter
\def\ps@IEEEtitlepagestyle{%
  \def\@oddfoot{\mycopyrightnotice}%
  \def\@oddhead{\hbox{}\@IEEEheaderstyle\leftmark\hfil\thepage}\relax
  \def\@evenhead{\@IEEEheaderstyle\thepage\hfil\leftmark\hbox{}}\relax
  \def\@evenfoot{}%
}
\def\mycopyrightnotice{%
  \begin{minipage}{\textwidth}
  \centering \scriptsize
  Copyright~\copyright~2024 IEEE. Personal use of this material is permitted. Permission from IEEE must be obtained for all other uses, in any current or future media, including\\reprinting/republishing this material for advertising or promotional purposes, creating new collective works, for resale or redistribution to servers or lists, or reuse of any copyrighted component of this work in other works.
  \end{minipage}
}
\makeatother
    
\begin{document}

\title{Sound Zone Control Robust To Sound Speed Change}

\author{
\IEEEauthorblockN{Sankha~Subhra~Bhattacharjee, Jesper~Rindom~Jensen, Mads~Græsbøll~Christensen}
\IEEEauthorblockA{\textit{Audio Analysis Lab}, \textit{Department of Electronic Systems}, \\ \textit{Aalborg University}, \textit{Denmark} \\ Email: \{ssbh; jrj; mgc\}@es.aau.dk}
}

\maketitle

\begin{abstract}
Sound zone control (SZC) implemented using static optimal filters is significantly affected by various perturbations in the acoustic environment, an important one being the fluctuation in the speed of sound, which is in turn influenced by changes in temperature and humidity (TH). This issue arises because control algorithms typically use pre-recorded, static impulse responses (IRs) to design the optimal control filters. The IRs, however, may change with time due to TH changes, which renders the derived control filters to become non-optimal. To address this challenge, we propose a straightforward model called sinc interpolation-compression/expansion-resampling (SICER), which adjusts the IRs to account for both sound speed reduction and increase. Using the proposed technique, IRs measured at a certain TH can be corrected for any TH change and control filters can be re-derived without the need of re-measuring the new IRs (which is impractical when SZC is deployed). We integrate the proposed SICER IR correction method with the recently introduced variable span trade-off (VAST) framework for SZC, and propose a SICER-corrected VAST method that is resilient to sound speed variations. Simulation studies show that the proposed SICER-corrected VAST approach significantly improves acoustic contrast and reduces signal distortion in the presence of sound speed changes.
\end{abstract}


\section{Introduction}
Sound zone control (SZC) has been applied to a wide range of applications, such as smart cars \cite{liao2017design, so2019subband, vindrola2021use}, home entertainment systems \cite{galvez2014personal}, controlling noise pollution in open concerts \cite{heuchel2018sound, heuchel2020large}, smartphones \cite{cheer2013practical} and neckbands \cite{jeon2020personal}. To achieve such spatial control of audio, several approaches have been proposed in the literature, the most popular approach being to pose SZC as a filter design problem, where the control filters are obtained by solving a suitable optimization problem. Commonly used control filter design methods include the acoustic contrast control (ACC) \cite{choi2002generation, jones2008personal, cai2013design, choi2015subband}, pressure matching (PM) \cite{poletti2008investigation, olivieri2017generation}, ACC-PM \cite{moller2012hybrid, chang2012sound, galvez2015time} and the recently proposed variable span trade-off (VAST) method which provides a continuum of control filter solutions corresponding to various degrees of trade-offs, with the ACC solution lying at one extreme and the PM solution at the other extreme \cite{lee2018unified, nielsen2018sound, shi2021generation, brunnstrom2022variable}. 

Traditionally, the different SZC methods use a pre-measured set of impulse responses (IRs), between all the loudspeakers and all the control points (sampled using microphones) in the bright zone (BZ) and the dark zone (DZ), to compute the control filters \cite{lee2018unified, galvez2015time}. Hence, in a practical implementation, for example in living rooms, car carbines, live concerts, etc., the control performance is highly sensitive to the accuracy between the pre-measured IRs and the true IRs \cite{betlehem2018temperature, zhang2022robust, zhang2023cgmm}. In other words, the computed control filters are optimal (depending on the cost function and rank parameter in VAST) for the acoustic characteristics and transducer locations at the time of IR measurement. However, after deployment of the SZC system, the acoustic characteristics may change, i.e., the pre-measured IRs may not exactly represent the current acoustic characteristics, causing the computed control filters to be sub-optimal, leading to degradation in SZC performance. Such changes of acoustic characteristics may occur due to various factors, such as variations in temperature and humidity (TH) causing change in sound propagation speed (SPS), change in transducer characteristics, or change in listener position \cite{elliott2012robustness, coleman2014acoustic, betlehem2018temperature, zhang2022robust_acc, zhang2022robust, zhang2023cgmm}. 

Change in SPS is one of the important factors leading to perturbations in IRs, and thus significantly impact SZC performance \cite{betlehem2018temperature, zhang2023cgmm, caviedes2019effect}. It is known from literature that an increase (or decrease) in SPS causes compressing (or stretching) of an IR along the time axis \cite{postma2016correction}. Using this fact, in this paper, we address the problem of SPS changes, which may be caused by temperature and/or humidity changes, in deploying SZC systems in indoor environments. First, we propose a new model to compensate IRs for SPS changes, called sinc interpolation-compression/expansion-resampling (SICER), which can compress or stretch an IR along time without introducing any discretization errors. Second, we integrate the SICER IR correction method into the popular VAST framework for SZC, and propose the SICER-corrected VAST method. The SICER-corrected VAST method can compute the optimal control filters after any temperature or humidity change in the acoustic environment without the requirement of re-measuring the new IRs, which is impractical after deploying the SZC system. Simulation studies carried out under reverberant conditions show significant improvement in acoustic contrast and reduction in signal distortion achieved by the proposed SICER-corrected VAST method, when changes in SPS occur.


\section{Sinc Interpolation - Compression/Expansion - Resampling (SICER)}
\label{section_sicer}
Here, we present the mathematical foundation to compute the new discrete time (DT) IR from an old DT IR, when SPS changes. Lets consider, we have a DT IR, $\bm{h} = [h(0), h(1), \cdots, h(N-1)]^{\textrm{T}} \in \mathbb{R}^{N \times 1}$, measured at a sound speed of $c_{\textrm{old}}$ (in m/s), where $n$ denotes the DT sample index. Now, under the assumption of ideal band-limited reconstruction, the continuous time (CT) reconstructed version of $h(n)$ can be written as
\begin{equation}
h_r(t) = \sum_{n=0}^{N-1} h(n) \textrm{sinc}(t - nT_s),
\end{equation}
where the subscript $r$ denotes the reconstructed version, and the limits of summation are determined by the starting and ending DT index of $h(n)$, $T_s$ denotes the sampling period (in seconds), with $f_s = \frac{1}{T_s}$ denoting the sampling frequency (in Hertz), and $\textrm{sinc}(t) = \frac{\sin(\pi t)}{\pi t}$ is the normalized sinc function. Now, lets consider the sound speed changes to $c_{\textrm{new}}$ and lets define a scaling factor $\beta = \frac{c_{\textrm{old}}}{c_{\textrm{new}}}$. If the SPS increases, i.e., $c_{\textrm{old}} < c_{\textrm{new}}$, we have $\beta<1 \Rightarrow \frac{1}{\beta} >1$. Similarly, when the sound speed decreases, $c_{\textrm{old}} > c_{\textrm{new}}$, we have $\beta>1 \Rightarrow \frac{1}{\beta} <1$. Now, when the SPS changes, we can apply a scaling of $\frac{1}{\beta}$ to the CT axis $t$ and the amplitude of $h_r(t)$, to estimate the new CT IR $h'(t)$ at sound speed $c_{\textrm{new}}$, as
\begin{equation}
h'(t) = h_r \bigg (\frac{t}{\beta} \bigg ) = \frac{1}{\beta} \sum_{n=0}^{N-1} h(n) \textrm{sinc} \bigg (\frac{t}{\beta} - nT_s \bigg ).
\label{eq_est_new_CT_IR}
\end{equation}
In (\ref{eq_est_new_CT_IR}), when the SPS increases, the scaling factor $\frac{1}{\beta} >1$ applies a compression along the time axis, whereas when the SPS decreases, the scaling factor $\frac{1}{\beta} <1$ applies stretching along the time axis. Moreover, in both the cases, the compression/stretching applied is proportional to the relative increase/decrease in SPS (via $\beta$). It may be noted that, the amplitude scaling of $\frac{1}{\beta}$ is necessary to ensure conservation of energy, i.e., the energy in the old IR and the new IR stay the same. 
We can now sample $h'(t)$ at the same sampling rate of $f_s = \frac{1}{T_s}$ by multiplying with an impulse train to obtain its DT version. Thus, using (\ref{eq_est_new_CT_IR}), the $m^{\textrm{th}}$ sample can be obtained as 
\begin{align}
h'(mT_s) &= h'(t) \delta(t - mT_s) \nonumber \\ &= \frac{1}{\beta} \Big [  \sum_{n=0}^{N-1} h(n) \textrm{sinc} \bigg (\frac{t}{\beta} - nT_s \bigg )  \Big ] \delta(t - mT_s) \nonumber \\ 
& = \frac{1}{\beta} \sum_{n=0}^{N-1} h(n) \textrm{sinc} \bigg ( \bigg (\frac{m}{\beta} - n \bigg ) T_s \bigg), \label{eq_sampled_est_of_new_IR}
\end{align}
where $m$ is a dummy variable denoting the DT sample index of $h'$. Dropping $T_s$ in (\ref{eq_sampled_est_of_new_IR}) and expressing the estimated DT IR at sound speed $c_{\textrm{new}}$ in terms of sample index $m$, we have
\begin{equation}
\begin{split}
h'(m) = \frac{1}{\beta} \sum_{n=0}^{N-1} h(n) \textrm{sinc} \bigg (\frac{m}{\beta} - n \bigg ) = \frac{1}{\beta} \bm{s}^{\mathrm{T}}(m) \bm{h},
\end{split}
\label{eq_DT_speed_change_IR_index_m}
\end{equation}
where $\bm{s}(m) = \Big [\textrm{sinc} \Big (\frac{m}{\beta} \Big ), \textrm{sinc} \Big (\frac{m}{\beta} - 1 \Big ), \cdots, \textrm{sinc} \Big (\frac{m}{\beta} - N + 1 \Big ) \Big ]^{\mathrm{T}}$, and $[\cdot]^{\mathrm{T}}$ is the transpose operator.
Consider $\bm{h}' \in \mathbb{R}^{M \times 1}$ to be of length $M$ in DT, then using (\ref{eq_DT_speed_change_IR_index_m}) all the DT samples of $\bm{h}'$ can be obtained as
\begin{align}
\bm{h}' &= \begin{bmatrix}  h'(0) \\ h'(1) \\ \vdots \\ h'(M-1) \end{bmatrix} = \frac{1}{\beta} \begin{bmatrix} \bm{s}^{\mathrm{T}}(0) \bm{h} \\ \bm{s}^{\mathrm{T}}(1) \bm{h} \\ \vdots \\ \bm{s}^{\mathrm{T}}(M-1) \bm{h} \end{bmatrix} = \frac{1}{\beta} \begin{bmatrix} \bm{s}^{\mathrm{T}}(0) \\ \bm{s}^{\mathrm{T}}(1) \\ \vdots \\ \bm{s}^{\mathrm{T}}(M-1) \end{bmatrix} \bm{h} \nonumber \\
&= \frac{1}{\beta} \Big [\bm{s}(0), \bm{s}(1), \cdots, \bm{s}(M-1) \Big ]^{\mathrm{T}} \bm{h} = \frac{1}{\beta} \bm{S}^{\mathrm{T}} \bm{h}, \label{eq_DT_speed_change_IR}
\end{align}
where $\bm{S} = \left [\bm{s}(0), \bm{s}(1), \cdots, \bm{s}(M-1) \right ] \in \mathbb{R}^{N \times M}$. Hence, we can estimate the DT IR $\bm{h}'$ at a new sound speed $c_{\textrm{new}}$, using a previously measured DT IR $\bm{h}$ at sound speed $c_{\textrm{old}}$ using (\ref{eq_DT_speed_change_IR}), given the knowledge of the sound speeds $c_{\textrm{old}}$ and $c_{\textrm{new}}$. We call (\ref{eq_DT_speed_change_IR}) the sinc interpolation-compression/expansion-resampling (SICER) model, as conceptually these are the sequence of steps involved in estimating $\bm{h}'$ from $\bm{h}$. 

\textbf{NOTE}: There is no explicit upsampling of $\bm{h}$ followed by downsampling, as in \cite{postma2016correction, prawda2023time}, and hence the SICER model in (\ref{eq_DT_speed_change_IR}) does not introduce any discretization errors. SPS can be obtained easily using TH readings from inexpensive commercial TH sensors \cite{cramer1993variation, sps_calculator}. Since we focus on indoor environments here, the affect of wind on SPS change is not considered. However, the proposed SICER model can be modified with wind models (for example, see Section 2.1 of \cite{caviedes2019effect}), given suitable wind sensors are available. Compression of an IR along time will consequently result in stretching along frequency in the frequency domain, and since the IRs exist in DT domain, this will lead to frequency domain aliasing effects in the new IR. In this paper, this aliasing is avoided by filtering the old IR to sufficiently attenuate high frequency contents close to the Nyquist frequency before applying SICER. Deriving the exact cut-off frequency of this low-pass filter is beyond the scope of this paper, and will be addressed in future work. Moreover, in future work, a theoretical explanation for the appearance of the amplitude scaling of $\frac{1}{\beta}$ in (\ref{eq_est_new_CT_IR}) will be explored. 

\section{SPS Change Robust Sound Zone Control}
\label{section_SZC_derivation}
Lets consider a simple static sound zone control (SZC) problem, where an array of $L$ loudspeakers are used to reproduce an audio signal in a so called bright zone (BZ) with minimum distortion, while keeping a nearby so called dark zone (DZ) as quiet as possible. Commonly, the control is achieved by considering the SZC problem as a filter design problem, where the input signal is pre-filtered using control filters $\bm{w}_l \in \mathbb{R}^{J \times 1}$, $l=1,2,\ldots, L$, i.e., one control filter corresponding to each loudspeaker. Let $K_b$ and $K_d$ represent the number of microphones used to sample the BZ and DZ, respectively. 
Let the DT IR from the $l^{\textrm{th}}$ loudspeaker to the $k^{\textrm{th}}$ microphone be $\bm{h}_{k,l} = [h_{k,l}(0), h_{k,l}(1), \cdots, h_{k,l}(N-1)]^{\textrm{T}} \in \mathbb{R}^{N \times 1}$, with $l=1,2,\ldots, L \land k=1,2,\ldots,K$, and $K=K_b$ for the BZ and $K=K_d$ for the DZ. 
To derive the optimal control filters, we consider the input signal to be a Kronecker delta function $\delta(n)$, since we consider a static SZC filter design \cite{galvez2015time, moles2020personal, moller2016sound, bhattacharjee2023study}. As our goal in this paper is to study the effect of and compensate for sound speed changes in SZC, considering the input signal to be $\delta(n)$ eliminates the effect of the input signal on the control filters \cite{galvez2015time, moles2020personal}, and allows us to study the effect of sound speed independently. The signal vector reproduced at the $k^{\textrm{th}}$ microphone in either BZ/DZ by all the loudspeakers, can be written as
\begin{equation}
\bm{y}_k^{(C)} = \sum_{l=1}^{L} h_{k,l}^{(C)} * w_l = \sum_{l=1}^{L} \bm{H}_{k,l}^{(C)} \bm{w}_l =  \bm{H}_{k}^{(C)} \bm{w},
\label{eq_kth_mic_signal_vec}
\end{equation}
where the superscript $C$ denotes the $\{\textrm{BZ},\textrm{DZ}\}$, $*$ is the DT convolution operator, $\bm{H}_{k,l} \in \mathbb{R}^{(N+J-1) \times J}$ is the convolution matrix formed from the DT IR $\bm{h}_{k,l}$, $\bm{H}_{k} = [\bm{H}_{k,1},  \cdots, \bm{H}_{k,L}] \in \mathbb{R}^{(N+J-1) \times LJ}$, $\bm{w} = [\bm{w}_1^{\textrm{T}},  \cdots, \bm{w}_L^{\textrm{T}}]^{\textrm{T}} \in \mathbb{R}^{LJ \times 1}$ and $\bm{y}_k^{(C)} \in \mathbb{R}^{(N+J-1)\times 1}$. Using (\ref{eq_kth_mic_signal_vec}), we can concatenate the individual signal vectors reproduced at all the $k=1,2,\ldots,K$ microphones, in either the BZ or DZ, as 
\begin{equation}
\bm{y}^{(C)} =  \left [\left (\bm{y}_1^{(C)} \right )^{\textrm{T}}, \left (\bm{y}_2^{(C)}\right )^{\textrm{T}} \cdots \left (\bm{y}_K^{(C)} \right )^{\textrm{T}} \right ]^{\textrm{T}} = \bm{H}^{(C)} \bm{w},
\label{eq_all_mic_signal_vec}
\end{equation}
where $\bm{H}^{(C)} = \Big [ \big ( \bm{H}_{1}^{(C)} \big )^{\textrm{T}}, \big ( \bm{H}_{2}^{(C)}\big )^{\textrm{T}}, \cdots, \big ( \bm{H}_{K}^{(C)}\big )^{\textrm{T}} \Big ]^{\textrm{T}} \\ \in \mathbb{R}^{K(N+J-1) \times LJ}$, with $K=K_b$ for $C=$ BZ and $K=K_d$ for $C=$ DZ. Let $\bm{d}_k$ be the desired signal to be reproduced at the $k^{\textrm{th}}$ microphone in BZ. In the chosen signal model, we can choose the $\bm{d}_k$ as the DT IR from a virtual loudspeaker, with the same vector dimensions as $\bm{y}_k$ in (\ref{eq_kth_mic_signal_vec}). Similar to (\ref{eq_all_mic_signal_vec}), the concatenated desired signal vector at all the microphones in the BZ can be written as $\bm{d} = [\bm{d}_1^{\textrm{T}}, \bm{d}_2^{\textrm{T}}, \cdots, \bm{d}_{K_b}^{\textrm{T}}]^{\textrm{T}} \in \mathbb{R}^{K_b(N+J-1) \times 1}$. The desired signal for the DZ is chosen as a vector of zeros, $\bm{0} \in \mathbb{R}^{K_d(N+J-1) \times 1}$, since it is intended to be as quiet as possible. Now, the control filter vector $\bm{w}$ can be obtained by minimizing the following cost function
\begin{equation}
\begin{split}
\zeta (\bm{w}) &= \| \bm{d} - \bm{y}^{(BZ)} \|^2  + \mu \| \bm{y}^{(DZ)} \|^2 \\
&= \bm{w}^{\textrm{T}} \bm{R}_b \bm{w} - 2 \bm{w}^{\textrm{T}} \bm{r}_b + \sigma_d^2 + \mu \bm{w}^{\textrm{T}} \bm{R}_d \bm{w},
\end{split}
\label{eq_cost_func}
\end{equation}
where $\| \cdot \|$ denotes the $\ell_2$-norm, $\bm{R}_b = \big (\bm{H}^{(\textrm{BZ})} \big)^{\textrm{T}} \bm{H}^{(\textrm{BZ})}$, $\bm{R}_d = \big (\bm{H}^{(\textrm{DZ})} \big)^{\textrm{T}} \bm{H}^{(\textrm{DZ})}$, $\bm{r}_b = \big (\bm{H}^{(\textrm{BZ})} \big)^{\textrm{T}} \bm{d}$, $\sigma_d^2 = \| \bm{d} \|^2$, $\mu$ controls the weightage between minimizing of reproduction error in BZ and minimizing the energy reproduced in DZ \cite{lee2020fast, zhang2023cgmm}. 

\subsection{VAST Approach}
To obtain the control filter, we consider the generalized framework of variable span linear filtering (VSLF), which includes several SZC methods like the 
ACC, PM,
and their variations as special cases \cite{nielsen2018sound, lee2018unified}. Following the signal model above, the control filter vector $\bm{w}$ lies in the space $\mathbb{R}^{LJ \times 1}$. We can consider a $V$-rank matrix $\bm{U}_V$, $1 \leq V \leq LJ$, formed by the set of basis vectors $\{\bm{u}_1, \bm{u}_2, \cdots, \bm{u}_V \}$ spanning a $V$ dimensional subspace of $\mathbb{R}^{LJ}$, as $\bm{U}_V = [\bm{u}_1, \bm{u}_2, \cdots, \bm{u}_V] \in \mathbb{R}^{LJ \times V}$. Then, a low-rank ($1 \leq V \leq LJ$) approximation of $\bm{w}$ can be written as $\bm{w} = \bm{U}_V \bm{a}$, where $\bm{a} \in \mathbb{R}^{V \times 1}$. Following the VSLF approach, the basis matrix $\bm{U}_V$ can be obtained by solving the generalized eigenvalue decomposition (GEVD) problem
\begin{equation}
\bm{U}_V^{\textrm{T}} \bm{R}_b \bm{U}_V = \bm{\Lambda}_V, \hspace{8mm} \bm{U}_V^{\textrm{T}} \bm{R}_d \bm{U}_V = \bm{I}_V,
\label{eq_joint_diag}
\end{equation}
where $\bm{\Lambda}_V = \textrm{diag}[\lambda_1, \lambda_2, \cdots, \lambda_V]$ contains the $V$ non-negative real-valued eigenvalues of $\bm{R}_d^{-1}\bm{R}_b$ in descending order $\lambda_1 \geq \lambda_2 \geq \cdots \geq \lambda_V$, $\bm{I}_V$ is the $V \times V$ identity matrix and $\textrm{diag}[\cdot]$ denotes a diagonal matrix formed by the elements of $[\cdot]$. Using $\bm{w} = \bm{U}_V \bm{a}$ in (\ref{eq_cost_func}), the control filter can be obtained as \cite{lee2018unified, lee2020fast}
\begin{equation}
\bm{w} = \sum_{v=1}^V \frac{\bm{u}_v^{\textrm{T}} \bm{r}_b}{\lambda_v + \mu} \bm{u}_v
\label{eq_optimal_weights_VAST}
\end{equation}
which is also known as the variable span trade-off (VAST) filter. 

\subsection{SICER-corrected VAST Approach}
In traditional SZC, the DT IRs $\bm{h}_{k,l}$, from each of the loudspeakers to each of the microphones are pre-measured, at a certain SPS. However, in practice, due to changes in TH, the SPS will change and consequently the true IRs also change \cite{olsen2017sound, betlehem2018temperature, postma2016correction, betlehem2015personal}. Hence, the pre-computed control filter in (\ref{eq_optimal_weights_VAST}) is no longer optimal, which causes degradation in control performance, as we will observe in Section \ref{section_simulation_study}. In such a scenario, using the SICER approach proposed in Section \ref{section_sicer}, the IRs can be corrected for changes in SPS. Lets consider we have a set of pre-measured DT IRs, $\bm{h}_{k,l}$, measured at a sound speed of $c_{\textrm{old}}$, $l=1,2, \hdots, L \land k=1,2,\hdots,K$, where $K=K_b$ for BZ and $K=K_d$ for DZ, using which the control filter vector in (\ref{eq_optimal_weights_VAST}) is derived. Now, lets consider the SPS changes to $c_{\textrm{new}}$, which can be computed from the application scenario, using available TH sensors \cite{cramer1993variation, sps_calculator}. Using the new sound speed $c_{\textrm{new}}$, the old sound speed $c_{\textrm{old}}$, and the old DT IRs $\bm{h}_{k,l}$, a new set of DT IRs $\bm{h}'_{k,l}$ at $c_{\textrm{new}}$ can be estimated using the SICER method in (\ref{eq_DT_speed_change_IR}). Next, following the VSLF framework, (\ref{eq_all_mic_signal_vec}) - (\ref{eq_optimal_weights_VAST}), a new control filter vector $\bm{w}'$ for sound speed $c_{\textrm{new}}$ can be derived, using the corrected DT IRs $\bm{h}'_{k,l}$. We call $\bm{w}'$ the SICER-corrected VAST control filter.

\section{Simulation Study}
\label{section_simulation_study}
We consider a SZC problem with one BZ and one DZ, similar to the setup in \cite{lee2020fast}, where the BZ and DZ are sampled with using $K_b = K_d = 37$ microphones each, with an inter-microphone spacing of $9$ cm and contains a linear loudspeaker array with $L=16$ loudspeakers, uniformly spaced by $6$ cm. All the loudspeakers and the control points lie on the x-y plane, at a height of $1.2$m. The $8^{\textrm{th}}$ loudspeaker is considered as the desired virtual source while computing the control filters. This SZC setup is placed in a shoe-box room of size $4.5$~m$\times 4.5~$m$\times 2.2$~m, and the IRs are simulated using the RIR Generator toolbox \cite{habets2006room, allen1979image}. IRs were generated with a reverberation time of $RT_{\textrm{60}} = 300$~ms, with IR length of $N=2,967$ and at a sampling frequency ($f_s = 16$~kHz). Further, we consider that the IRs at a SPS of $c=343$ m/s (old IRs) are available for deriving the SZC control filters, while the SZC is deployed in perturbed TH conditions with a SPS of: (Scenario I) $c=333$ m/s and (Scenario II) $c=353$ m/s, denoting a reduction and increase in SPS, respectively. The old and new DT IRs, i.e., $\bm{h}$ and $\bm{h}'$ are considered to have the same length $N=M$. We assume that when deployed, the SZC system contains a TH sensor to provide temperature and humidity readings, from which the SPS $c$ can be easily computed \cite{cramer1993variation, sps_calculator}. The trade-off parameter, $\mu=1$, is considered for VAST, with a control filter length, $J=800$. The input signal during SZC deployment is a clean speech signal of duration $11.2$ seconds, obtained by concatenating $4$ speech segments from male and female speakers from the NOIZEUS database \cite{loizou2013speech} after resampling to $16$ kHz. Studies with a $\delta(n)$ input signal and also another filter length, $J$, were also carried out, which lead to the same observations, hence are not presented here to avoid redundancy. 

For comparison, the performance metrics considered are: (a) Time domain acoustic contrast (TD AC), (b) Time domain normalized signal distortion (TD nSDP), (c) Frequency domain acoustic contrast (FD AC), and (d) Frequency domain normalized signal distortion (FD nSDP), defined as \cite{lee2020fast, shi2021generation}:
\begin{align}
&\text{TD AC}  = \Big [ K_d\left (\bm{y}^{(BZ)} \right )^{\textrm{T}} \bm{y}^{(BZ)}\Big ] \Big [K_b \left (\bm{y}^{(DZ)} \right )^{\textrm{T}} \bm{y}^{(DZ)} \Big ]^{-1} \label{eq_TD_AC} \\
&\text{TD nSDP} = \Big [\| \bm{d} - \bm{y}^{(BZ)} \|^2\Big ] \Big [\| \bm{d} \|^2 \Big ]^{-1} \label{eq_TD_nSDP} \\
&\text{FD AC}(f) = \Big [K_d \sum_{k=1}^{K_b} \vert Y_k^{(BZ)}(f) \vert^2 \Big ] \Big [ K_b \sum_{k=1}^{K_d} \vert Y_k^{(DZ)}(f) \vert^2 \Big]^{-1}  \label{eq_FD_AC} \\
&\text{FD nSDP}(f) = \Big [ \sum_{k=1}^{K_b} \vert   Y_k^{(BZ)}(f) - D_k(f) \vert^2 \Big ] \Big[ \sum_{k=1}^{K_b} \vert   D_k(f) \vert^2 \Big]^{-1} \label{eq_FD_nSDP} 
\end{align}
where $Y_k^{(BZ)} = \mathcal{F} \left \{ \bm{y}_k^{(BZ)} \right \}$ and $D_k = \mathcal{F} \left \{\bm{d}_k \right \}$ for $k=1,2,\hdots, K_b$, $Y_k^{(DZ)} = \mathcal{F} \left \{\bm{y}_k^{(DZ)} \right \}$ for $k=1,2,\hdots, K_d$, $\mathcal{F}\{\cdot\}$ denotes Fourier transform and $f$ denotes the frequency bin. 
%
%
Moreover, a higher AC and a lower nSDP is desirable \cite{lee2020fast}. In the results presented here, `GT' represents the case where control filters are computed using the ground truth (GT) IRs at the actual SPS; `NC', i.e., no correction, represents the case where control filters are computed using the IRs at $c=343$ m/s, i.e. the old IRs; and `SICER' represents the case where SICER-corrected IRs are used to compute the control filters. While deploying the SZC, the performance is tested using the GT IRs (true IRs at $c=333$ m/s in Scenario I and $c=353$ m/s in Scenario II) for all the three cases. 

\begin{figure}[b]
    \centering
    \includegraphics[width=\linewidth]{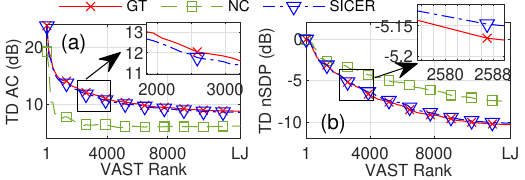}
    \caption{Scenario I: Comparison of (a) TD AC and (b) TD nSDP, across different VAST ranks. $LJ = 12800$.}
    \label{fig_TD_AC_nSDP_343_333_down_J_800_b_3}
\end{figure}


\begin{figure}[]
    \centering
    \includegraphics[width=0.95\linewidth]{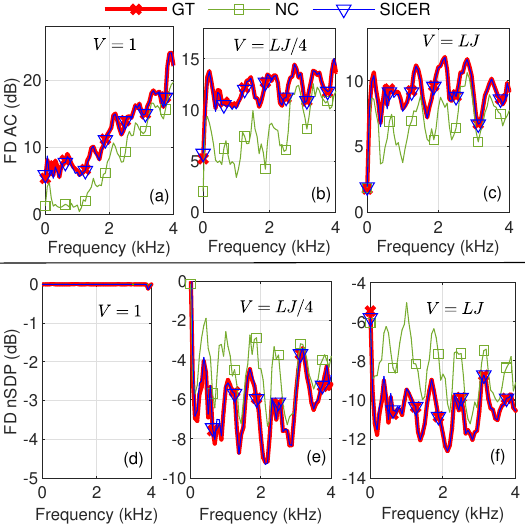}
    \caption{Scenario I: Comparison of (a-c) FD AC and (d-f) FD nSDP, for VAST ranks (a,d) $V=1$ (ACC), (b,e) $V=LJ/4$ and (c,f) $V=LJ$ (PM).}
    \label{fig_speed_red_speech_T60_300_J_800}
\end{figure}

\begin{figure}[]
    \centering
    \includegraphics[width=\linewidth]{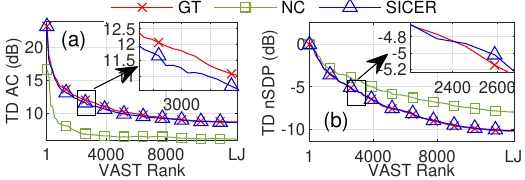}
    \caption{Scenario II: Comparison of (a) TD AC and (b) TD nSDP, across different VAST ranks. $LJ = 12800$.}
    \label{fig_TD_AC_nSDP_343_353_up_J_800_b_3}
\end{figure}


\begin{figure}[]
    \centering
    \includegraphics[width=0.95\linewidth]{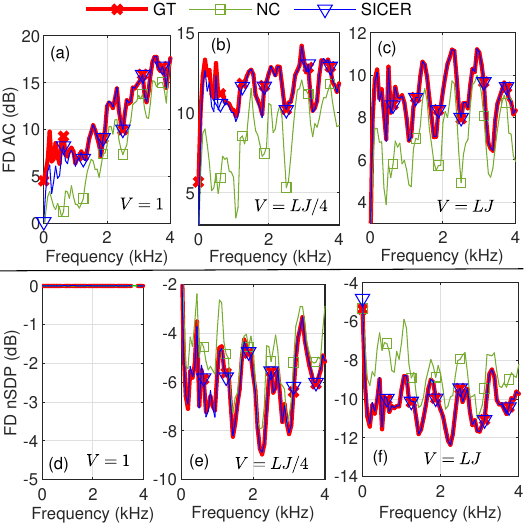}
    \caption{Scenario II: Comparison of (a-c) FD AC and (d-f) FD nSDP, for VAST ranks (a,d) $V=1$ (ACC), (b,e) $V=LJ/4$ and (c,f) $V=LJ$ (PM).}
    \label{fig_speed_inc_speech_T60_300_J_800}
\end{figure}

\subsection{Scenario I: Sound Propagation Speed Reduction To $c=333$ m/s}
\label{subsection_szc_speed_reduction}
For this scenario, Fig.\ \ref{fig_TD_AC_nSDP_343_333_down_J_800_b_3} shows the comparison of the time domain AC and nSDP achieved in the three different cases of `GT', `NC' and `SICER', for $100$ integer values of VAST ranks uniformly sampled in the interval $\left [ 1, LJ \right ]$, where rank $1$ corresponds to the ACC solution and rank $LJ$ corresponds to the PM solution. From Fig.\ \ref{fig_TD_AC_nSDP_343_333_down_J_800_b_3}(a), we can see that the TD AC reduces when the control filter solution computed using IRs at $c=343$ m/s is used to control the sound in the zones at $c=333$ m/s. On the other hand, when the control filters are computed using the SICER-corrected IRs, the achieved TD AC increases and gets close to the `GT'. A similar observation can be made in terms of nSDP from Fig.\ \ref{fig_TD_AC_nSDP_343_333_down_J_800_b_3}(b) where we see that, the nSDP increases (i.e., performance degrades) when the control filters computed using IRs at $c=343$ m/s is used to control sound in the zones at $c=333$ m/s, whereas the control filter computed using SICER-corrected IRs achieves nSDP close to `GT'.
Fig.\ \ref{fig_speed_red_speech_T60_300_J_800} shows the frequency domain AC and nSDP comparison for the VAST ranks $V=1$ (ACC), $V=LJ/4$ and $V=LJ$ (PM).
From Fig.\ \ref{fig_speed_red_speech_T60_300_J_800}, similar to the TD comparison, we can observe that using the control filter solution computed using IRs at $c=343$ m/s to control sound at $c=333$ m/s, results in a performance drop both in terms of AC and nSDP. On the other hand, using the control filter computed using the SICER-corrected IRs to control sound at $c=333$ m/s improves the performance and gets close to the `GT' case. 

\subsection{Scenario II: Sound Propagation Speed Increase To $c=353$ m/s}
\label{subsection_szc_speed_increase}
Fig.\ \ref{fig_TD_AC_nSDP_343_353_up_J_800_b_3} shows the comparison of the TD AC and nSDP for this scenario, for the three cases `GT', `NC' and `SICER', for $100$ uniformly sampled integer values of VAST ranks $\in \left [ 1, LJ \right ]$. From Fig.\ \ref{fig_TD_AC_nSDP_343_353_up_J_800_b_3}(a), we can observe that the AC reduces when the control filter solution computed using IRs at $c = 343$ m/s is used for control at $c = 353$ m/s. By instead using SICER correction on the IRs, the corresponding control filters increase the AC and restores performance close to the `GT' case. Similarly, from the nSDP comparison in Fig.\ \ref{fig_TD_AC_nSDP_343_353_up_J_800_b_3}(b), we can see that the performance drops when no IR correction is done after sound speed change, whereas by using the SICER correction, the performance is restored close to `GT'. The performance comparison in the frequency domain for VAST ranks $\{1, LJ/4, LJ\}$ is shown in Fig.\ \ref{fig_speed_inc_speech_T60_300_J_800}, where, again, we can observe the improvement in performance achieved by using SICER-corrected VAST compared to no correction after sound speed increase. 

It may be noted that the results presented in this section represent ideal conditions, since the IRs are noise-free and the sound speed is constant along the entire propagation path from the loudspeakers to the microphones (i.e., same TH at all points in the room) at a certain time. However, in a practical scenario, this may seldom be true and thus some performance degradation can be expected when using the proposed SICER-corrected VAST method. Nevertheless, the results presented here validates the effectiveness of the proposed SICER-corrected VAST method and the SZC performance achieved can be considered to represent the upper bound of improvement achievable, which makes it quite promising for practical scenarios. Evaluation for practical measured IRs will be carried out in future studies. 

\section{Conclusion}
In this paper, we propose a new approach to compensate for the affect of sound speed changes on sound zone control (SZC) and improve its robustness. An IR correction method called sinc interpolation-compression/expansion-resampling (SICER) is first proposed, which compensates IRs for sound speed changes. The SICER model is then incorporated into the recently proposed SZC framework of variable span linear filtering, and the SICER-corrected VAST approach is proposed for SZC robust to sound speed changes (resulting from temperature and/or humidity changes). SZC results using simulated IRs under reverberant conditions show that the proposed SICER-corrected VAST method can significantly improve SZC performance in terms of acoustic contrast and signal distortion, which gets close to the true optimal filter performance, for both sound speed reduction and increase. 


\newpage
\bibliographystyle{IEEEtran}
\bibliography{IEEEabrv_v_1_14, stringss_1}

\end{document}